# Enhanced optical nonlinearities in epitaxial quantum dots lasers on silicon for future photonic integrated systems

Jianan Duan,[1,2,*] Weng W. Chow,[3] Bozhang Dong,[1] Heming Huang,[1] Songtao Liu,[4] Justin C. Norman,[4] John E. Bowers,[4] and Frédéric Grillot[1,5]

[1] LTCI, Télécom Paris, Institut Polytechnique de Paris, 91120, Palaiseau, France
[2] State Key Laboratory on Tunable Laser Technology, School of Electronic and Information Engineering, Harbin Institute of Technology, Shenzhen, 518055, China
[3] Sandia National Laboratories, Albuquerque, NM 87185, USA
[4] Department of Electrical and Computer Engineering, Materials Department, University of California Santa Barbara, Santa Barbara, CA 93106, USA
[5] Center for High Technology Materials, University of New-Mexico, Albuquerque, NM 87106, USA

**Abstract**. Four-wave mixing (FWM) is an important nonlinear optical phenomenon that underlines many of the discoveries and device applications since the laser was invented. Examples include parametric amplification, mode-locked pulses and frequency combs, and in the quantum optics regime, entangled-photon generation, squeezed-state production and optical transduction from the visible to infrared wavelengths. For quantum dot systems, the basic understanding of FWM is limited by the conventional investigation method, which concentrates on the FWM susceptibility measured with optical amplifiers. This paper addresses this weakness by performing laser experiments to account for all optical nonlinearities contributing to the FWM signal. Meanwhile, we gain valuable insight into the intricate interplay among optical nonlinearities. Using quantum dot lasers directly grown on silicon, we achieved FWM conversion efficiency sufficient to demonstrate self-mode-locking in a single-section laser diode, with sub-ps mode-locked pulse duration and kHz frequency-comb linewidth. A comparison with first-principles based multimode laser theory indicates measured FWM conversion efficiencies that are close to the theoretical limit. An advantage over earlier studies and crucial to confidence in the results are the quality and reproducibility of state-of-the-art quantum dot lasers. They make possible the detailed study of conversion efficiency over a broad parameter space, and the identification of the importance of p-doping. Systematic improvement based on our understanding of underlying physics will lead to transform limited performance and effective compensation of intrinsic and extrinsic effects, such as linewidth enhancement and background dispersion. The integration of FWM with lasing impacts numerous optoelectronic components used in telecom and datacom.

*Jianan Duan, E-mail: jianan.duan@hotmail.com

## 1 Introduction

Photonic integrated circuits (PICs) on silicon can significantly advance the level of component integration and performance necessary for taking both conventional and quantum information processing outside the laboratory[1]. The advantages of silicon-based PICs are the availability of manufacturing approaches using modern nanofabrication techniques as well as the potential for miniaturization and integration of optoelectronic components with complementary functionalities[2]. In this situation, quantum dot (QD) nanostructures are highly promising semiconductor atoms that can be integrated either monolithically or heterogeneously on a compact and scalable platform[3-6]. As a direct consequence of the size-confinement effect of the trapped electrons and holes, QD based photonic devices have shown remarkable properties. In particular, epitaxial QD lasers directly grown on silicon have recently led to record performance such as ultra-low threshold currents, high temperature continuous-wave operation, very long device lifetimes as well as high yield and much better scalability[3]. In addition, the use of p-doping significantly improves their thermal stability and reliability[7]. It also reduces the linewidth enhancement factor (α-factor), resulting in reflection insensitivity, which is of vital importance for isolator-free PICs[8]. Extending these advances to integrated photonics technologies will lead to silicon platforms with on-chip nonclassical light sources, large versatile photonic logic, quantum information storage, and highly efficient detectors[9].

Four-wave mixing (FWM) is useful in optical communications for all-optical signal processing and for wavelength-division multiplexing (WDM) systems, which are a key component in coherent communication technologies[10-12]. FWM is known to drive the phase and mode-locking properties observed in comb QD lasers[13,14]. Therefore, our interest in FWM susceptibility involves QD lasers, where mode locking is possible with both single- and multi-section diode lasers[13,15]. The interest comes from applications in telecom and datacom. For example, in the case of WDM systems, a single mode locked laser producing a frequency comb can potentially replace the large number of lasers presently necessary for the task. A single-section mode-locked laser using self-mode locking amplifies the advantages even further. However, there are serious challenges because with self-mode locking, the gain medium alone has to produce the multimode lasing that leads to broad emission bandwidth, and the FWM contributes to the locking of the frequencies. Within the inhomogeneously broadened distribution of QDs, the optical nonlinearities of light-matter interactions give rise to both mechanisms. To control self-mode locking to the extent that it can be employed in applications, such as WDM, requires a deeper understanding than we have presently of the intricate interplay of physics associated with mode competition and FWM. Our laser experiments and accompanying theoretical analysis are designed with exactly that goal in mind.

Within a nonlinear gain medium that has a third-order nonlinear susceptibility, the beating between two co-polarized fields at different frequencies results in the occurrence of wave-mixing and the generation of two new fields. The highly nonlinear optical fibers are certainly good candidates for achieving efficient wave-mixing conversion, however, the required interaction length of several meters together with the large pump power makes them not suitable for monolithic integration[16,17]. To overcome this limit, efficient FWM with relatively low power consumption can also be achieved with micro-ring resonators, however, at the price of implementing sophisticated low-loss bus waveguide designs[18-20]. In addition, high fabrication costs may be another issue compared with more compact devices, such as semiconductor optical amplifier (SOA) or semiconductor laser sources. In the latter case, the FWM is essentially driven by the carrier density pulsation (CDP) which reinforces the wave-mixing conversion efficiency, nevertheless, the nanosecond timescale of the carrier recombination lifetime leads to a slow response speed[21]. In this context, as compared to bulk and quantum well (QW) semiconductors, QDs exhibit larger optical nonlinearities with faster response speed. The QD gain material is spectrally broader, and the fast carrier dynamics along with the lower linewidth enhancement factor improves the conversion efficiency[22]. Prior work concentrated on QD SOAs[23,24] and lasers[25,26] grown on lattice matched substrates. In the former, although the higher conversion efficiency can be achieved through the larger linear gain in SOAs, the inherently stronger amplified spontaneous emission noise limits the optical signal-to-noise ratio.

This paper reports recent results from a study aim at understanding FWM in QD lasers epitaxially grown on silicon, taking into account their enhanced cavity resonances and reduced amplified spontaneous emission noise. In our investigation, we measured and analyzed the nonlinear optical contributions resulting in mode competition, gain saturation, carrier-induced refractive index, α-factor and creation of combination tones, all of which have roles in self-mode-locking. The experiments were performed on epitaxial QD lasers with inhomogeneous broadening below 10 meV, similar to ones that we used to achieve frequency combs with RF linewidth less than 100 kHz and mode-locked pulse below 500 fs[13]. The present motivation is to build on prior results and to confirm that the contributing nonlinearities in general combine to provide a strong mechanism of self-mode-locking. Success will significantly reduce complexity and energy requirement in WDM systems.

## 2   Device description

The Fabry-Perot (FP) QD laser material was grown a 300 mm on-axis (001) GaP/Si substrate. The active region includes five periods of QD layers. The dot-in-a-well QD layer composed of InGaAs QWs asymmetrically encompassing the InAs dots with a 2 nm prelayer below and a 5 nm capping layer on the top. Each QD layer is separated by a 37.5 nm GaAs spacer. For p-doped QD lasers, a 10 nm p-GaAs layer at a target hole concentration of $5\times10^{17}$ cm$^{-3}$ (10 extra holes per QD) is sandwiched between a 10 nm undoped GaAs layer and a 17.5 nm undoped GaAs layer. It is noted that the gain is temperature sensitive in QD laser due to the thermal spreading of holes. To solve this problem, the p-doping in the GaAs spacer can counter the influence of closely spaced whole energy level hence the ground state transition of QD is full of holes. The inclusion of p-doping brings many advantages for laser devices. On the one hand, the p-doping can ease the thermal spread of holes and lead to rather temperature-insensitive characteristics such as threshold current, α-factor, relative intensity noise and optical feedback resistance[10,27,28]. On the other hand, it can also eliminate gain saturation and gain broadening, hence improving the high-frequency response of QD lasers[29,30]. Furthermore, the optimized growth conditions contribute to a narrow photoluminescence full-width-at-half-maximum below 30 meV, which transforms an inhomogeneous broadening width of 10 meV[8]. The output facet has a facet coating of 60% power reflectivity while the rear facet has a value of 90%.

## 3 Four-wave mixing experiment

Fig. 1 depicts the FWM experimental setup with optical injection locking configuration. Two narrow linewidth tunable lasers are used as drive laser and probe laser, the light of which is incorporated by a 90/10 coupler and then injected into the QD laser using optical circulator and lens-end fiber. The drive laser is used to lock the gain peak mode of the FP modes while the probe laser is used to generate the FWM with the locked FP modes. The polarization controllers are applied to align the polarization of two tunable lasers with QD laser for realizing the maximum conversion. The FWM optical spectrum is recorded from the optical circulator by an optical spectrum analyzer (OSA) with a 20 pm resolution. The working temperature of the QD laser is kept at 293 K throughout the experiment using a thermoelectric cooler.

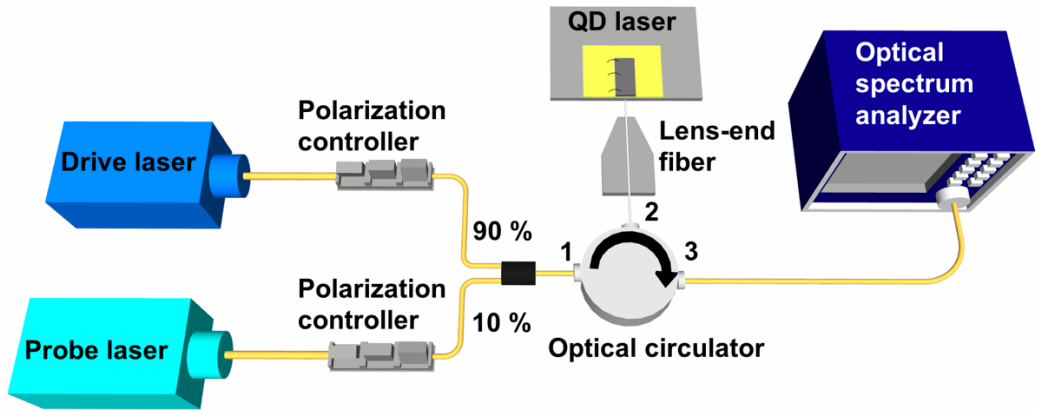

FIG. 1. Optical injection locking setup used for the four-wave mixing experiments.

The probe-drive injection frequency detuning Δ is defined as the frequency difference between the drive laser and probe laser. The drive laser is used to lock the longitudinal FP mode at the gain peak while the side modes are deeply suppressed, hence generating the drive signal for wave-mixing. Within the stable-locked regime, the frequency of the probe laser is then tuned to have the probe signal coincide with one of the suppressed FP cavity modes to obtain maximum conversion. Fig. 2(a) and 2(b) show typical FWM spectra recorded for undoped laser with Δ=114 GHz and for p-doped laser with Δ=89 GHz. It is noted that the free spectral range (FSR) is 38 GHz for the undoped laser and 30 GHz for the p-doped one. The middle peak represents the FP mode locked in the QD gain peak at the drive laser frequency with deeply suppressed sidemodes. The left peak is the probe signal mode while the right one is the wave-mixing induced converted signal. The different colored curves in Figs 2(a) and (b) show the signal power increase with increasing the probe power.

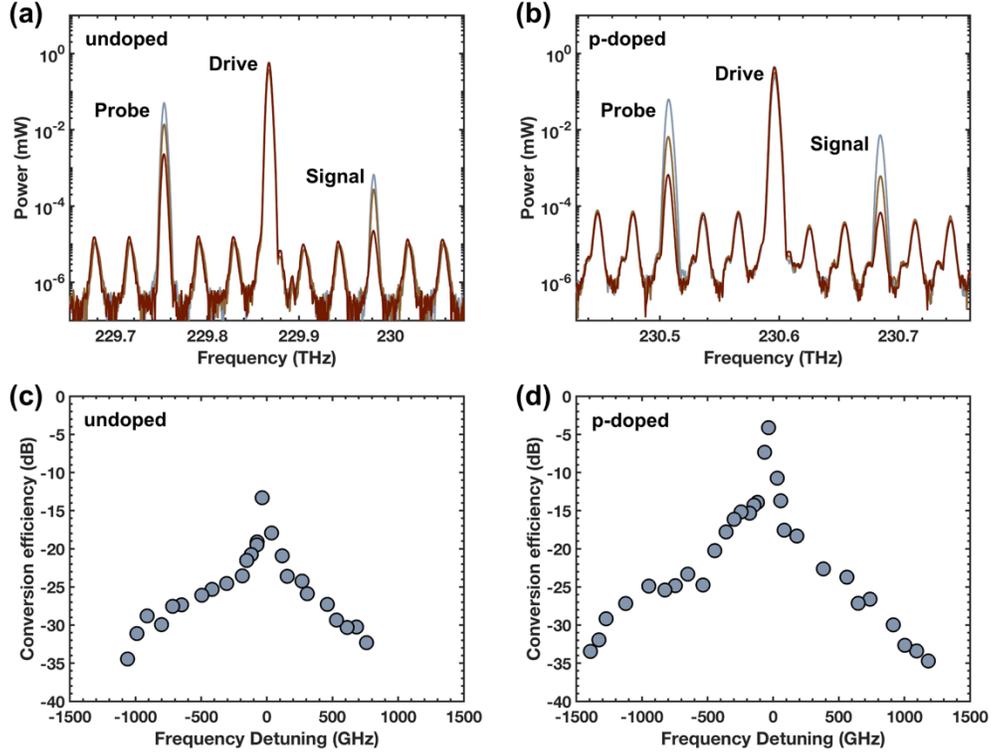

FIG. 2. Optical spectra from a four-wave mixing experiment for (a) undoped laser with up-conversion frequency detuning of 114 GHz and (b) p-doped laser with up-conversion frequency detuning of 89 GHz (probe-drive mode number difference $\Delta m = 3$). The different colored lines indicate signal power increase with increasing the probe power. Conversion efficiency of four-wave mixing for (c) undoped laser and (d) p-doped laser as a function of the probe-drive frequency detuning.

The conversion efficiency is expressed as[22]:

$$\eta_{CE} = \frac{P_{Signal}}{P_{Probe}} \qquad (1)$$

with $P_{Signal}$ is the optical power of the converted signal and $P_{Probe}$ is the probe signal power injected into the laser. In the experiment, these powers are obtained from the measured optical spectrum. The laser-fiber coupling loss is estimated by calculating the ratio between the laser free-space output power and the laser power coupled in the lens-end fiber. The total losses include coupling loss and fiber loss, which are considered in the spectra in order not to over-estimate the value of $\eta_{CE}$. The value of $\eta_{CE}$ is expressed in logarithmic scale (in dB) in this paper.

In the first set of measurements, we keep the probe power constant, adjust the probe frequency by multiples of the FSR and record the resulting maximal $\eta_{CE}$ for each FSR. Fig. 2(c) and 2(d) compare the measured $\eta_{CE}$ for undoped and p-doped lasers as a function of the probe-drive injection frequency detuning. The up-conversion (down-conversion respectively) refers to the converted signal has a frequency higher (lower respectively) than the drive, hence the probe frequency minus drive frequency is negative (positive respectively). As shown, the probe-drive injection frequency detunings at which $\eta_{CE}$ is measured are multiples of the FSR of the FP laser and the corresponding $\eta_{CE}$ are the local maxima. The maximal $\eta_{CE}$ is found at -13 dB for the undoped laser and -4 dB for the p-doped laser when the probe laser injects into the first longitudinal mode next to the drive signal. These results are consistent with native InAs/GaAs QD laser while the maximal $\eta_{CE}$ is a bit larger[22]. For the undoped laser, the $\eta_{CE}$ is kept above -35 dB for up-conversion with frequency detunings up to 1 THz and up to 760 GHz for down-conversion. The frequency detunings for $\eta_{CE}$ above -35 dB in p-doped laser are larger than that of the undoped laser, which are up to 1.4 THz for up-conversion and 1.2 THz for down-conversion. The larger maximal $\eta_{CE}$ and frequency detuning in p-doped laser are due to the increased material gain because the p-doping eliminates gain saturation and gain broadening[31]. It is noted that the static conversion at low-frequency region is determined by the carrier density pulsation (CDP), which is directly caused by the probe-drive beating. For larger frequency detunings, carrier heating (CH) and spectral hole burning (SHB) become the dominant mechanisms occurring within sub-picosecond timescales. Although $\eta_{CE}$ are higher at low frequency hence increasing the third-order nonlinear susceptibility, the very large bandwidth provided by the CH and SHB remain very promising for broadband wavelength conversion. The different bandwidth between up and down conversion is attributed to the asymmetry in the gain profile leading to a non-zero α-factor. In theory, with a zero α -factor, the bandwidths are perfectly symmetric. Although the α-factor is small in epitaxial QD lasers, destructive interferences still persist due to the phase condition arising between the different nonlinear processes (eg. CDP, CH and SHB). In addition, the profiles between up-conversion and down-conversion are found to be more symmetric in p-doped laser than that of undoped one, which is due to lower α-factor. As previously demonstrated, the α-factor of the p-doped laser is found to be as low as 0.13 at the gain peak, which is lower than that of the undoped laser with α-factor of 0.3[27].

## 4 Theory

Fig. 3(a) shows a drawing of the laser showing the QD active region, which consists of 5 layers of InGaAs QWs embedding InAs QDs and separated by GaAs barriers. A transverse waveguide is formed by AlGaAs cladding layers. For a FWM experiment, the injected intensities are $I_d^{inj}$ and $I_p^{inj}$, the measured output intensities are $I_d^{out}$, $I_p^{out}$ and $I_s^{out}$. The model computes the corresponding intracavity intensities $I_d$, $I_p$ and $I_s$. Intensities in the other modes are $I_n$. Fig. 3(b) demonstrates the electronic states contributing to the laser transitions. The electronic states considered in the model are the ground and excited states from the QDs, and the continuum of states from the QWs. Laser transitions involve only the inhomogeneously broadened QD ground states. The laser modes participating in an intracavity FWM experiment are shown in Fig. 3(c).

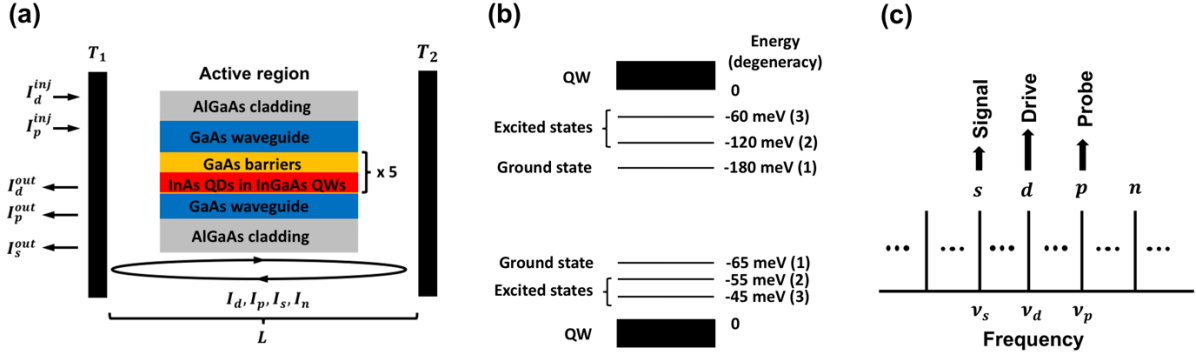

FIG. 3. (a) Sketch of laser showing QD active region inside optical cavity of length *L* and facet transmissions $T_1$ and $T_2$. The active region consists 5 layers of InGaAs QWs embedding InAs QDs and separated by GaAs barriers. (b) Electronic states considered in the model are the ground and excited states from the QDs, and the continuum of states from the QWs. (c) Laser modes participating in an intracavity four-wave mixing experiment. Injections are at the drive and probe modes, labeled *d* and *p*, respectively. The signal appears in the mode labeled *s*. The model also tracks other lasing or nonlasing modes, collectively labeled as *n*.

To extract the FWM coefficient from the measured data, we developed a laser model for the intensity in cavity mode, as sketched in Fig.3(c). A semiclassical treatment[32] is chosen to treat gain saturation, mode competition and multi-wave mixing on equal footing[33].

From the semiclassical derivation, the intracavity drive, probe and signal intensities evolve according to:

$$\frac{d}{dt}i_d = \left(g_d^{sat} - \frac{\nu_0}{Q}\right)i_d + \frac{c}{2Ln_b}\sqrt{T_1 i_d^{inj} i_d} + s \quad (2)$$

$$\frac{d}{dt}i_p = \left(g_p^{sat} - \frac{\nu_0}{Q}\right)i_d + \frac{c}{2Ln_b}\sqrt{T_1 i_p^{inj} i_p} + s \quad (3)$$

$$\frac{d}{dt}i_s = \left(g_s^{sat} - \frac{\nu_0}{Q}\right)i_s + \sqrt{2}|\theta_{sdpd}|i_d\sqrt{i_s i_p} + s \quad (4)$$

Additionally, the derivation gives for the other lasing and non-lasing modes:

$$\frac{d}{dt}i_n = \left(g_n^{sat} - \frac{\nu_0}{Q}\right)i_s + s \quad (5)$$

In the above equations, we define a dimensionless intensity:

$$i_j = \frac{2}{\varepsilon_0 c n_B}\left(\frac{\wp}{2\hbar\gamma}\right)^2 I_j \quad (6)$$

where $\varepsilon_0$ and *c* are the permittivity and speed of light in vacuum, $n_B$ is the background refractive index, $\wp$ is the QD dipole matrix element, $\gamma$ is the dephasing rate and $I_m$ is the intensity in the $m^{th}$ cavity mode. Also in the equations, the cavity linewidth *L* is expressed in terms of an average cavity mode frequency $\nu_0$ and the cavity quality factor *Q*:

$$\frac{\nu_0}{Q} = \frac{c}{n_b}\left[\alpha_{abs} - \frac{1}{2L}\ln(R_1 R_2)\right] \quad (7)$$

where $\alpha_{abs}$ is the intracavity absorption, $R_1 = 1-T_1$ and $R_2 = 1-T_2$. Each cavity mode experiences a saturated modal gain given by:

$$g_n^{sat} = \frac{g_n}{1+\kappa_{nn}i_n+\sum_{m\neq n}\kappa_{n,m}i_m} \tag{8}$$

where $g_n$ is the unsaturated (small signal) modal gain, $\kappa_{nn}$ is the self-gain compression factor and $\kappa_{n,m}$ is the cross gain compression factor. To account for spontaneous we add $s$ to Eqs. (2) - (5):

$$s = \frac{\hbar v_0 \beta B_{2d}}{2\varepsilon_0 n_B^2 d_{qw}}\left(\frac{\wp}{2\hbar\gamma}\right)^2 n_{e0}n_{h0} \tag{9}$$

where $\beta$ is the spontaneous emission factor, $B_{2d}$ is the bimolecular carrier recombination coefficient, $d_{qw}$ is the thickness of each QW layer embedding the QDs, $n_{e0}$ and $n_{h0}$ are the carrier densities in the QD electron and hole ground states.

The gain parameters $g_n$, $\kappa_{nn}$ and $\kappa_{n,m}$ depend on carrier densities. We group the electron and hole densities into those populating the inhomogeneously broadened QD ground states $n_{e0}$ and $n_{h0}$, those populating the QD excited states $n_{e1}$ and $n_{h1}$, and those populating the QW layers $n_{e2}$ and $n_{h2}$. The semiclassical laser theory also provides the following equations of motion for the carrier densities:

$$\frac{dn_{e0}}{dt} = -\frac{\varepsilon_0 n_b^2 d_{qw}}{\hbar v_0 \Gamma_{conf}}\left(\frac{2\hbar\gamma}{\wp}\right)^2 \sum_n g_n^{sat} i_n - \gamma_r\left(n_{e0} - n_{e0}^{eq}\right) - B_{2d}n_{e0}n_{h0} - \gamma_{nr}n_{e0} - C_{2d}n_{e0}^2 n_{h0} \tag{10}$$

$$\frac{dn_{h0}}{dt} = -\frac{\varepsilon_0 n_b^2 d_{qw}}{\hbar v_0 \Gamma_{conf}}\left(\frac{2\hbar\gamma}{\wp}\right)^2 \sum_n g_n^{sat} i_n - \gamma_r\left(n_{h0} - n_{h0}^{eq}\right) - B_{2d}n_{e0}n_{h0} - \gamma_{nr}n_{h0} - C_{2d}n_{e0}n_{h0}^2 \tag{11}$$

$$\frac{dn_{e1}}{dt} = -\gamma_r\left(n_{e1} - n_{e1}^{eq}\right) - B_{2d}n_{e1}n_{h1} - \gamma_{nr}n_{e1} - C_{2d}n_{e1}^2 n_{h1} \tag{12}$$

$$\frac{dn_{h1}}{dt} = -\gamma_r\left(n_{h1} - n_{h1}^{eq}\right) - B_{2d}n_{e1}n_{h1} - \gamma_{nr}n_{h1} - C_{2d}n_{e1}n_{h1}^2 \tag{13}$$

$$\frac{dn_{e2}}{dt} = \frac{\eta J}{e} - \gamma_r\left(n_{e2} - n_{e2}^{eq}\right) - B_{2d}n_{e2}n_{h2} - \gamma_{nr}n_{e2} - C_{2d}n_{e2}^2 n_{h2} \tag{14}$$

$$\frac{dn_{h2}}{dt} = \frac{\eta J}{e} - \gamma_r\left(n_{h2} - n_{e2}^{eq}\right) - B_{2d}n_{e2}n_{h2} - \gamma_{nr}n_{h2} - C_{2d}n_{e2}n_{h2}^2 \tag{15}$$

The first terms on the right-hand side of Eqs. (10) and (11) are from stimulated emission, where $\Gamma_{conf}$ is the mode confinement factor. The terms with an effective carrier scattering rate $\gamma_r$ approximate carrier relaxation of all the states towards quasi-equilibrium. The quasi-equilibrium densities $n_\alpha^{eq}$ are determined by the conservation of carrier population. We assume equilibrium within each density group. However, in general, $n_\alpha \neq n_\alpha^{eq}$ because of dynamic population bottleneck. Details on implementing this phenomenological approach to modeling carrier scattering effects and its accuracy compared to quantum kinetic calculations are discussed in earlier publications[34,35]. Also in the carrier density equations are the carrier loss contributions. The terms with $\gamma_{nr}$ and $C_{2d}$ account for carrier losses from defect and Auger scattering, respectively. Lastly, in Eqs. (14) and (15), $J$ is the injected current density and $\eta$ is the injection efficiency from electrodes to QW states.

The saturated gain in Eq. (8) is determined using the following expressions. Starting with the small signal gain, semiclassical laser theory gives:

$$g_n = [f_{e0,n} + f_{h0,n} - 1]\frac{v_0 \wp^2 N_{qd}\Gamma_{conf}}{\varepsilon_0 n_b^2 \hbar\gamma d_{qw}}\Lambda_n \quad (16)$$

where

$$\Lambda_n = \int_{-\infty}^{\infty} d\omega \frac{1}{\sqrt{2\pi}\Delta_{inh}} \exp\left[-\left(\frac{\omega-\omega_0}{\sqrt{2}\Delta_{inh}}\right)^2\right] L(\omega - \nu_n) \quad (17)$$

$$L(\omega - \nu_n) = \frac{\gamma^2}{\gamma^2 + (\omega - \nu_n)^2} \quad (18)$$

and the quantity inside the square bracket is the population inversion, assuming equilibrium among QD ground state populations within the inhomogeneously broadened distribution of width $\Delta_{inh}$. The quantities $f_{e0,n}$ and $f_{h0,n}$ are Fermi functions evaluated for the QD populations contributing to the lasing transition at frequency $\nu_n$, based on the instantaneous QD ground state densities $n_{e0}$ and $n_{h0}$.

Semiclassical laser theory also gives the self- and cross- gain compression factors:

$$\kappa_n = 3\frac{\gamma}{\gamma_{ab}}\frac{1}{\Lambda_n}\int_{-\infty}^{\infty} d\omega \frac{1}{\sqrt{2\pi}\Delta_{inh}} \exp\left[-\left(\frac{\omega-\omega_0}{\sqrt{2}\Delta_{inh}}\right)^2\right] L^2(\omega - \nu_n) \quad (19)$$

$$\kappa_{n,m} = \frac{\gamma}{\gamma_{ab}}\left(1 + \frac{\zeta_{nm}}{2}\right)\frac{1}{\Lambda_n}\int_{-\infty}^{\infty} d\omega \frac{1}{\sqrt{2\pi}\Delta_{inh}} \exp\left[-\left(\frac{\omega-\omega_0}{\sqrt{2}\Delta_{inh}}\right)^2\right]\{2L(\omega-\nu_n)L(\omega-\nu_m) +$$
$$Re\{D_{\gamma_{ab}}(\nu_m - \nu_n)D_\gamma(\omega - \nu_n)[D_\gamma(\nu_m - \omega) + D_\gamma(\nu_n - \omega)]\}\} \quad (20)$$

where

$$D_\alpha(\chi) = \frac{\alpha}{\alpha + i\chi} \quad (21)$$

and $\zeta_{nm}$ accounts for the effects of spatial hole burning in the presence of carrier diffusion. In the limit of no carrier diffusion,

$$\zeta_{nm} = \frac{1}{L}\int_0^L dz \cos[2(k_n - k_m)z] = \delta_{n,m} \quad (22)$$

If carrier diffusion dominates, $\zeta_{nm} = 1$ for all combinations of $n$ and $m$. The focus on our investigation is on the relative phase angle term[32] in Eq. (4), which contributes to the signal intensity via FWM.

To extract the FWM susceptibility for mode locking $\chi^{(3)}_{sdpd}$, we solve Eqs. (2) - (5) and Eqs. (10) - (15), with the relative phasing angle coefficient $\theta_{sdpd}$ as an input parameter. The value of $\theta_{sdpd}$ that best fit the experimental data then gives the susceptibility:

$$\chi^{(3)}_{sdpd} = \frac{\sqrt{2}n_b}{v_0 \Gamma_{conf}}\left(\frac{\wp}{2\hbar\gamma}\right)^2 |\theta_{sdpd}| \quad (23)$$

In an earlier study,[36,37] a first-principle based multimode laser theory was used to derive the following expression for the relative phase angle coefficient:

$$|\theta_{sdpd}| = (f_{e,s} + f_{h,s} - 1)\frac{v_0 \wp^2 N_{QD}^{(2d)}\Gamma_{conf}}{2\varepsilon_0 n_b^2 \hbar\gamma d_{qw}}\frac{\gamma}{\gamma_{ab}}\frac{1}{2}(1 + 2\zeta)$$

$$\times \int_{-\infty}^{\infty} d\omega \frac{1}{\sqrt{2\pi}\Delta_{inh}} e^{-\left[\frac{\omega-\omega_0}{\sqrt{2}\Delta_{inh}}\right]^2}$$
$$\times \left|D_\gamma(\omega + \nu_p - 2\nu_d)D_{\gamma_{ab}}(\nu_p - \nu_d)\left[D_\gamma(\nu_p - \omega) + D_\gamma(\omega - \nu_d)\right]\right| \quad (24)$$

A quantity that is often extracted from experiments is $\chi^{(3)}_{sdpd}$ divided by the material small signal gain in inverse length[25,38]. One can see the attractiveness by using Eqs. (16), (23) and (24) to obtain:

$$\xi \equiv \chi^{(3)}_{sdpd} \frac{c\Gamma_{conf}}{n_b g_s}$$

$$= \frac{\sqrt{2}c}{\nu_0 \Lambda_s}\left(\frac{\wp}{2\hbar\gamma}\right)^2 \frac{\gamma}{\gamma_{ab}} \frac{1}{2}(+2\zeta) \times \int_{-\infty}^{\infty} d\omega \frac{1}{\sqrt{2\pi}\Delta_{inh}} e^{-[(\omega-\omega_0)/\sqrt{2}\Delta_{inh}]^2}$$
$$\times \left|D_\gamma(\omega + \nu_p - 2\nu_d)D_{\gamma_{ab}}(\nu_p - \nu_d)\left[D_\gamma(\nu_p - \omega) + D_\gamma(\omega - \nu_d)\right]\right| \quad (25)$$

The results show a quantity that depends only on the electronic structure and broadenings associated with carrier scattering. Its generality arises from being independent of the laser configuration, in terms of confinement factor, QD density, heterostructure layer thicknesses and injection current.

TABLE I. Device parameters. † Computed for QW embedding QDs.

| Device parameter | Symbol | Value |
|---|---|---|
| QD layers | $n_{qd}$ | 5 |
| QD density | $N_{qd}$ | $4\times10^{14}\ m^{-2}$ |
| QD layer thickness | $d_{qw}$ | 7 nm |
| Barrier thickness | $d_b$ | 40 nm |
| Waveguide cross section | $w \times d_{wg}$ | 4.9 μm × 0.3 μm |
| Cavity length | $L$ | 1.1 mm |
| Facet reflections | $R_1, R_2$ | 0.6, 0.9 |
| Mode spacing | $\Delta_c$ | 38 GHz undoped<br>30 GHz p-doped |
| Inhomogeneous width | $\Delta_{inh}$ | 10 meV |
| Mode confinement factor | $\Gamma_{conf}$ | 0.06 † |

TABLE II. Model input parameters. † Based on $C_{2d} = C_{3d}/d_{qw}^2$ with $C_{3d} = 1.7\times10^{-38}\ m^6 s^{-1}$.

| Model parameter | Symbol | Value |
|---|---|---|
| Dephasing rate | $\gamma$ | $10^{13}\ s^{-1}$ |
| QD-QW scattering rate | $\gamma_r$ | $10^{13}\ s^{-1}$ |

| Inter-QD scattering rate | $\gamma_{ab}$ | $5\times10^{11}$ $s^{-1}$ |
| --- | --- | --- |
| Defect loss rate | $\gamma_{SRH}$ | $4\times10^{8}$ $s^{-1}$ undoped <br> $2\times10^{9}$ $s^{-1}$ p-doped |
| Bimolecular recombination coefficient | $B_{2d}$ | $1.4\times10^{-8}$ $m^2s^{-1}$ |
| Auger coefficient † | $C_{2d}$ | $3.5\times10^{-22}$ $m^4s^{-1}$ |
| Spontaneous emission factor | $\beta$ | $10^{-3}$ |
| Intracavity absorption | $\alpha_{abs}$ | 1200 $m^{-1}$ undoped <br> 1600 $m^{-1}$ p-doped |
| Spatial hole burning | $\xi$ | 0.5 |

Before using the model to extract $\chi^{(3)}_{sdpd}$, we need to determine the values for its input parameters. First are the device parameters determined through fabrication. These parameters are listed in Table I and their values are entered directly into the model. Then come the model parameters that are in Table II. From the list, the spontaneous emission and Auger coefficients $B_{2d}$ and $C_{2d}$, respectively, are from published data[39,40]. The reminders are determined by anchoring computed laser behavior to measured ones, as discussed below.

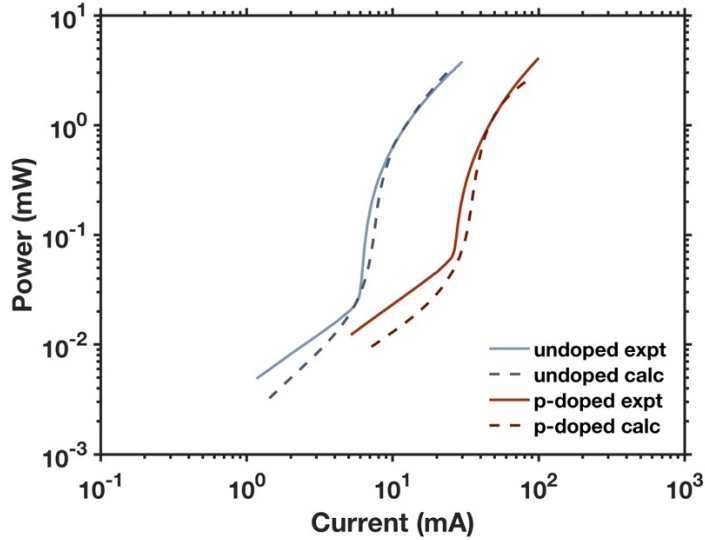

FIG. 4. Output power versus current for undoped and p-doped lasers used in four-wave mixing experiments. The solid lines are the experimental results and the dashed lines are the simulation. Input parameters for the calculated curves are given in Table II.

Fig. 4 presents the output power as a function of bias current of both undoped and p-doped QD lasers used in the study. The undoped laser has threshold current of 6 mA at 293 K, while the p-doped laser has a bit larger threshold current of 27 mA due to the high free carrier absorption caused by the large number of holes. The dashed lines are from the theoretical calculation using the device and gain medium parameters listed in Tables I and II. Both lasers have ground state emission at 1.3 μm. The fitting of power-current curves pins the intracavity absorption $\alpha_{abs}$, injection efficiency $\eta$ and defect (SRH) loss rate $\gamma_{nr}$. We note that the higher intracavity absorption and defect loss rate with p-doping is consistent with previous studies on p-doping effects[31]. The p-doping changes linear and nonlinear gain properties through state filling and carrier scattering induced dephasing, as well as alters the $\gamma_{nr}$ and $\alpha_{abs}$[31]. These effects are taken into account in the model.

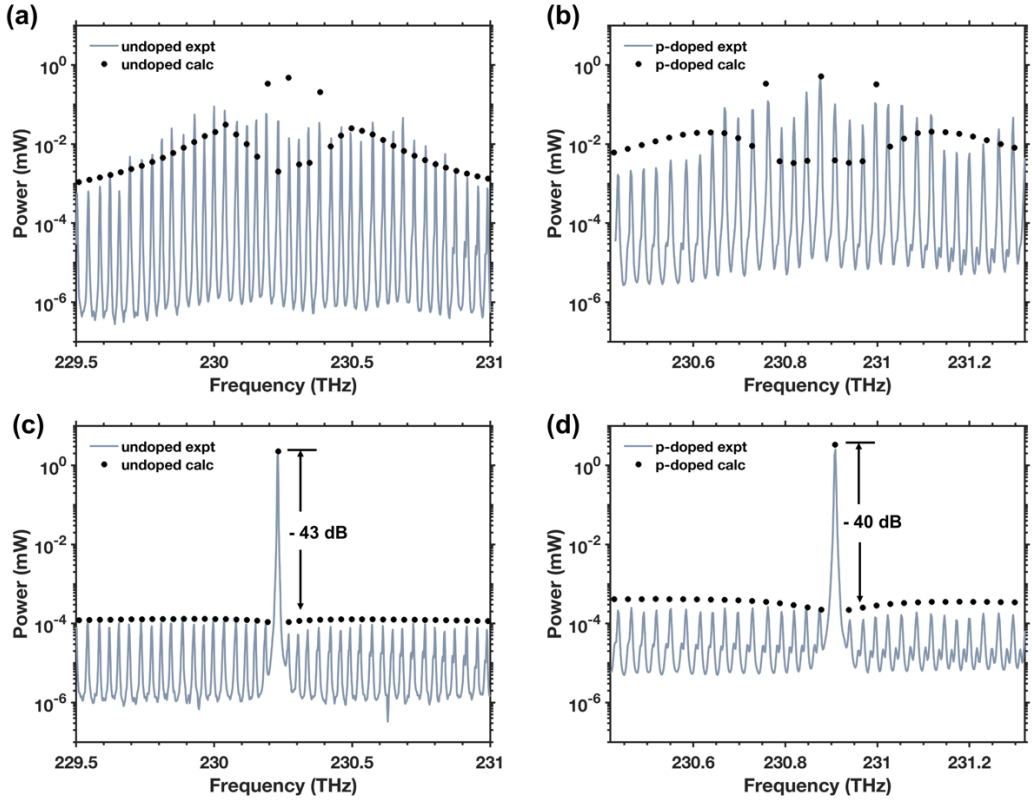

FIG. 5. Measured optical spectra from undoped (a) free-running and (c) injection-locked lasers operating at twice threshold current. Measured optical spectra from p-doped (b) free-running and (d) injection-locked lasers operating at twice threshold current. The points are the calculated mode power.

Optical spectra of both lasers under free running and injection-locked operating at twice threshold current are shown in Fig. 5. When fitting to the free-running lasing spectra in Fig. 5(a) and 5(b), we observed a transition from a smooth multimode frequency comb to a spikey one with increasing current. As the transition is caused by mode competition, we believe that it is more important to reproduce this excitation dependent transition than the actual spectral shape. Doing so fixes the scattering rates $\gamma$, $\gamma_r$ and $\gamma_{ab}$. The chosen values for $\gamma$ and $\gamma_r$ are consistent with those from quantum kinetic calculations[41]. The spectral fit also specifies the spatial hole burning parameter. We found that a value of $\zeta = 0.5$, midway between maximal and negligible carrier diffusion effects, to provide reasonable description of the measurements.

The final confirmation of the input parameters is from reproducing the spectra for the injection-locked lasers without any further adjustment of parameters. As shown in the Figs. 5(c) and 5(d), the model correctly describes the 43 dB and 40 dB side mode suppression for the undoped and p-doped lasers, respectively.

## 5 Extracting four-wave mixing coefficient

Due to their fast carrier-carrier and carrier-phonon scatterings, QDs have shown large optical nonlinearities and the fast FWM conversion has been achieved in QD SOAs as a result of fast carrier scattering induced deeper spectral holes[31]. The optical nonlinearities of epitaxial QD lasers on silicon are now analyzed based on a microscopic level model containing quantum mechanical electron-hole polarization[36,37]. To do so, the net FWM coefficient (normalized to the material gain) $\xi \equiv \chi^{(3)}_{sdpd}/g_s$ is extracted from probe-drive laser experiment and connected to multimode semiclassical laser theory by considering the gain competition. We repeat the measurement giving the spectra in Figs. 2(a) and 2(b) for a range of probe power and for three different frequency detuning between drive and probe fields. The points in Figs. 6 summarize the experimental results and show the measured signal power with increasing probe power for fixed drive power. The plotted solid curves are obtained from the model with the relative phase angle coefficient $\theta_{sdpd}$ as a fitting parameter, which is chosen to bound the experimental points. The model simulations are in good agreement with the experiment results. For up to 4 mode number separation Δm between probe and drive fields in both undoped laser and p-doped lasers, we found the experimental points to be bounded by the choice of $\theta_{sdpd}$ giving $\xi \equiv \chi^{(3)}_{sdpd}/g_s$ between 4 and $8\times10^{-21}$ $m^3V^{-2}$. With higher separations, the extracted $\xi$ falls below $4\times10^{-21}$ $m^3V^{-2}$.

Within the variance in the measured data, both undoped and p-doped lasers have basically the same values for $\xi$. However, as shown in Fig.6, the gain in signal power with increasing probe power (slope of data points and curves) indicates higher net FWM gain with p-doping. These results emphasize the need to consider gain competition and $\chi^{(3)}_{sdpd}$ on equal footing and under similar experimental conditions when evaluating mode-locking performance. Hence, caution should be exercised when drawing conclusions using only $\chi^{(3)}_{sdpd}$ measured in amplifier experiments. Rather, measurements should be made directly with lasers with configurations closely resembling the lasers one is attempting to mode lock. Along with fitting to experimental data, we also computed the conversion efficiency between the signal and drive power using a first-principle based multimode laser theory. It gives the relative phase angle coefficient as described in Eq. (24). The dashed curves show the theoretical prediction for the conversion efficiency as a function of ratio between probe and drive power. Comparison of theory and experiment indicates that the FWM susceptibility from our samples is consistent with the theoretical predictions.

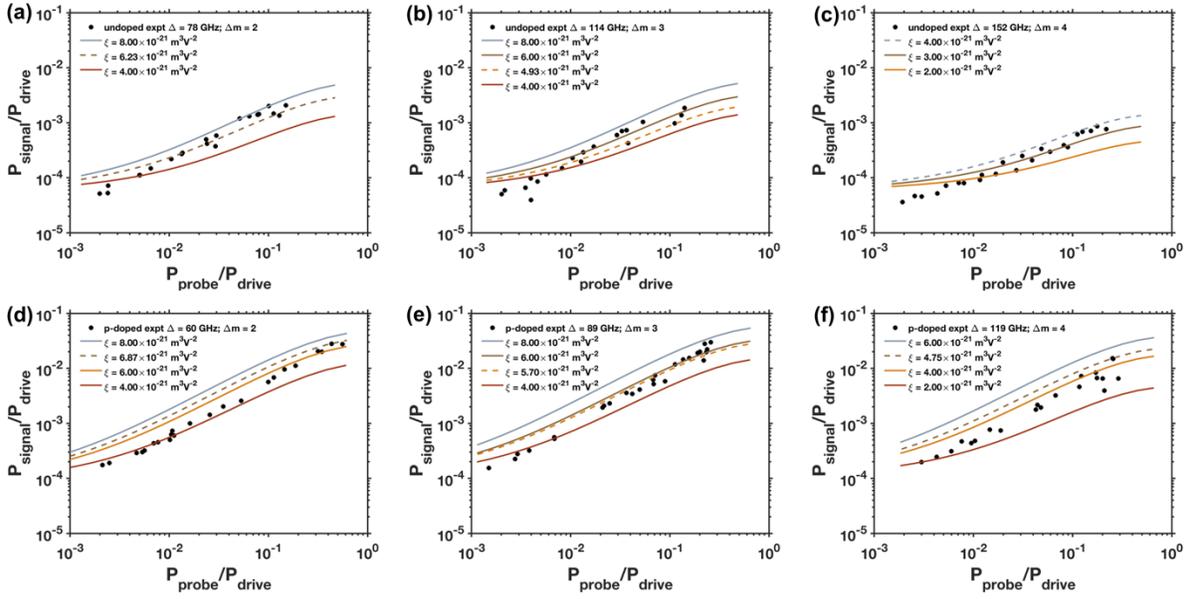

FIG. 6. Signal power versus probe power for (a, b, c) undoped and (d, e, f) p-doped lasers operating at twice threshold current. Both powers are normalized to the drive power. The data points are from experiment with probe-drive injection frequency detuning $\Delta$ as indicated. The calculated curves in (a, d) are for probe-drive mode number difference $\Delta m = 2$, that in (b, e) are for probe-drive mode number difference $\Delta m = 3$, and that in (c, f) are for probe-drive mode number difference $\Delta m = 4$. The dashed curves are computed using $\xi = \chi^{(3)}/g$ calculated from multimode laser theory. The other curves use $\xi$ is an input parameter.

## 5 Conclusions and perspectives

In conclusion, we investigated the nonlinear optical properties in semiconductor QD lasers directly grown on silicon. Our experiments show strong optical nonlinearities that allow demonstration of self-mode-locking in lasers fabricated for use in silicon-based PICs. Gain saturation, mode competition and multi-wave mixing are connected through the active region optical nonlinearities, with the leading contribution arising from the third-order electron-hole polarization. In contrast to amplifier experiments, the probe-drive laser measurements provide valuable insight to the intricate interplay of optical nonlinearities during device operation. We show that the gain in signal power with increasing the probe power produces a much higher net FWM gain, the latter being even further magnified owing to p-doping. However, when the FWM gain is no longer enough to overcome the mode competition, the signal power drops. On the top of that, at large FWM susceptibility, the FWM gain can exceed the cavity losses hence lasing by FWM can take place. Consequently, the laser experiments provide clear differences from the amplifier ones because of the gain saturation by the drive and probe intracavity fields.

While epitaxial QD lasers on silicon are already strong building blocks of on-chip integrated quantum photonic circuits[42,43], further analysis could possibly extend this work to semiconductor-based quantum information systems. For instance, the high third-order nonlinear susceptibility can be used for light squeezing to reduce the noise below the standard quantum limit[44,45]. Squeezed states can be generated by using a FWM source made with epitaxial QDs[46]. Overall, these results allow us to present the new insights of the third order non-linearity for the mode locking mechanism of the QDs comb laser, which is crucial for classical coherent communication. The experiments are solid and quantitatively compared to the microscopic model. It could also pave the way for the control of optical nonlinearities in epitaxial QDs at the crossway between classical and quantum physics in the future.

## Acknowledgements

This work is supported in part by the Center for Integrated Nanotechnologies (CINT) through a Rapid Access proposal (2019BRA0032). Authors also acknowledge the financial support of the DARPA PIPES program, the Institut Mines-Télécom, and the Research Startup Fund of HITSZ.

## Conflict of interests

The authors declare no conflict of interest.